\documentclass[twocolumn,nofootinbib,floatfix,amsmath,amssymb,showpacs,superscriptaddress]{revtex4}

\usepackage{rotating}
\usepackage{graphicx}
\usepackage{dcolumn}
\usepackage{bm}
\usepackage{color}
\usepackage[bordercolor=white,backgroundcolor=gray!30,linecolor=black,colorinlistoftodos]{todonotes}
\usepackage[pdftex,bookmarks=true]{hyperref}
\hypersetup{
    colorlinks,
    citecolor=black,
    filecolor=black,
    linkcolor=black,
    urlcolor=black
}

\newcommand{\sgc}{\Sigma_c}
\newcommand{\omc}{\Omega_c}
\newcommand{\occ}{\Omega_{cc}}
\newcommand{\xcc}{\Xi_{cc}}

\newcommand{\ffac}[2]{F_{#1,{\cal #2}}}

\begin{document}

\title{Electromagnetic structure of charmed baryons in Lattice QCD}

\author{K. U. Can}
\author{G. Erkol}
\author{B. Isildak}
\affiliation{Department of Natural and Mathematical Sciences, Faculty of Engineering, Ozyegin University, Nisantepe Mah. Orman Sok. No:13, Alemdag 34794 Cekmekoy, Istanbul Turkey}
\author{M. Oka}%
\affiliation{Department of Physics, H-27, Tokyo Institute of Technology, Meguro, Tokyo 152-8551 Japan}
\author{T. T. Takahashi}
\affiliation{Gunma National College of Technology, Maebashi, Gunma 371‐8530, Japan }

\date{\today}

\begin{abstract}
As a continuation of our recent work on the electromagnetic properties of the doubly charmed $\xcc$ baryon, we compute the charge radii and the magnetic moments of the singly charmed $\sgc$, $\omc$ and the doubly charmed $\occ$ baryons in 2+1 flavor Lattice QCD. In general, the charmed baryons are found to be compact as compared to the proton. The charm quark acts to decrease the size of the baryons to smaller values. We discuss the mechanism behind the dependence of the charge radii on the light valence- and sea-quark masses. The magnetic moments are found to be almost stable with respect to changing quark mass. We investigate the individual quark sector contributions to the charge radii and the magnetic moments. The magnetic moments of the singly charmed baryons are found to be dominantly determined by the light quark and the role of the charm quark is significantly enhanced for the doubly charmed baryons. 

\end{abstract}
\maketitle

\section{Introduction}

Electromagnetic form factors play an important role in describing the internal structure of hadrons. They reveal valuable information about the size and the shape of the hadrons. Obviously, determining these form factors is an important step in our understanding of the hadron properties in terms of quark-gluon degrees of freedom.

Last two decades have witnessed an enormous experimental and theoretical endeavor, which has been concentrated in particular on the form factors of pion and nucleon, that are the lightest QCD bound states. There has also been an increasing activity in determining the electromagnetic structure of octet mesons and baryons. The theoretical challenge is to understand these quantities from QCD. In the framework of Lattice QCD --the only known method that starts directly from QCD-- the electromagnetic form factors have been extensively studied. Lattice computations have now reached an advanced level so that the electromagnetic structure of the nucleon can be probed for pion masses as low as~$m_\pi\sim 180$~MeV~\cite{Collins:2011mk, Alexandrou:2011db}.  

From this perspective, one intriguing question is how the structure of the hadrons gets modified in the heavy-quark regime, like in the case of charm hadrons. While there exist experimental results for the light baryons revealing their spectrum and electromagnetic properties, only the spectrum of the charmed baryons are accessible by experiments for the time being. Future charm factories like BES-III and PANDA at GSI are expected to probe the charm sector. 

Recently we have extracted the electromagnetic form factors of the doubly charmed $\xcc(ccu/ccd)$ baryons on a 2+1 flavor $32^3 \times 64$ lattice with a lattice spacing of 0.096~fm. We have computed the form factors up to $\sim$1.5~GeV$^2$ and using these we have extracted relevant quantities such as the electric and magnetic charge radii and the magnetic moment of the baryon. We have found that due to the heavy $c$ quark the $\xcc$ has much smaller electric and magnetic charge radii as compared to, \emph{e.g.}, the proton. Indeed, such a compactness may be an indication of a peculiar quark distribution inside the baryon~\cite{Brodsky:2011zs}.

One puzzling property of the $\xcc$ baryon as reported by the SELEX Collaboration is that its mass isospin splitting is much larger than that of any other hadron. Using the Cottingham formula Brodsky \emph{et al.} showed that the large isospin splitting is an implication for the compact structure of $\xcc$~\cite{Brodsky:2011zs}. Therefore, our findings about the size of the doubly heavy baryon is in accordance with the experimental results.

The main conclusion that can be drawn from our lattice analysis in Refs.~\cite{Can:2012tx, Can:2013zpa} is that the charmed hadrons are compact. One question that has remained unanswered is how the charge radii change as the extra light quark in the composition of the baryon gets heavier. If the effect of the extra light quark is to decrease the string tension between the two-charm component~\cite{Yamamoto:2007nn}, we shall expect an increase in the size of the hadron as this extra quark gets heavier. This can be further tested by changing the extra $u$/$d$ light quark, say, with the $s$ quark, which recalls a study of the charmed baryons with strangeness. Yet as appealing is the electromagnetic structure of the singly charmed baryons, which will provide a broader perspective to inner dynamics of heavy baryons and complete the picture.  

In this work, we extend our previous lattice analysis on the doubly charmed $\Xi_{cc}$ baryon so as to include the singly charmed $\Sigma^{(0,++)}_c(cuu,cdd)$, $\Omega^{0}_c(css)$ baryons and the doubly charmed $\Omega^{+}_{cc}(ccs)$ baryon. In particular we compute the electric and magnetic charge radii, and the magnetic moments of these baryons. Since we have run our simulations on the same lattice setup as we have used in our previous works, we refer the reader to Refs.~\cite{Can:2012tx, Can:2013zpa} for details. For the heavy quark, we employ the Fermilab method. We shall give the lattice parameters pertaining to the $s$ quark below.

\section{Theoretical Formalism}\label{sec2}

\subsection{Lattice formulation}
Electromagnetic form factors can be calculated by considering the baryon matrix elements of the electromagnetic vector current $V_\mu=  \sum \limits_{q}{} e_q \bar{q}(x) \gamma_{\mu} q(x)$, where $q$ runs over the quark content of the given baryon. The matrix element can be written in the following form
	\begin{align}
		\begin{split}\label{matel}
	\langle {\cal B}(p)|V_\mu|{\cal B}(p^\prime)\rangle= \bar{u}(p) &\left[\gamma_\mu F_{1,{\cal B}}(q^2) \right. \\
	&\left. +i \frac{\sigma_{\mu\nu} q^\nu}{2m_{\cal B}} F_{2,{\cal B}}(q^2) \right]u(p),
	\end{split}
	\end{align}
where $q_\mu=p_\mu^\prime-p_\mu$ is the transferred four\--momentum. Here $u(p)$ denotes the Dirac spinor for the baryon with four-momentum $p^\mu$ and mass $m_{\cal B}$. The Sachs form factors $F_{1,{\cal B}}(q^2)$ and $F_{2,{\cal B}}(q^2)$ are related to the electric and magnetic form factors by
\begin{align}
	G_{E,\cal{B}}(q^2)=\ffac{1}{B}(q^2)+\frac{q^2}{4m_{\cal B}^2}\ffac{2}{B}(q^2),\\
	G_{M,\cal{B}}(q^2)=\ffac{1}{B}(q^2)+\ffac{2}{B}(q^2).
\end{align}

\begin{widetext}
Our method of computing the matrix element in Eq.~\eqref{matel}, which was employed to extract the nucleon electromagnetic form factor, follows closely that of Ref.\cite{Alexandrou:2011db}. Using the following ratio
\begin{align}
\begin{split}\label{ratio}
	&R(t_2,t_1;{\bf p}^\prime,{\bf p};\mathbf{\Gamma};\mu)=
	\cfrac{\langle F^{{\cal B V_\mu B^\prime}}(t_2,t_1; {\bf p}^\prime, {\bf p};\mathbf{\Gamma})\rangle}{\langle F^{{\cal B}{\cal B}}(t_2; {\bf p}^\prime;\Gamma_4)\rangle}\left[\frac{\langle F^{{\cal BB}}(t_2-t_1; {\bf p};\Gamma_4)\rangle \langle F^{{\cal BB}}(t_1; {\bf p}^\prime;\Gamma_4)\rangle \langle F^{{\cal BB}}(t_2; {\bf p}^\prime;\Gamma_4)\rangle}{\langle F^{{\cal BB}}(t_2-t_1; {\bf p}^\prime;\Gamma_4)\rangle\langle F^{{\cal BB}}(t_1; {\bf p};\Gamma_4)\rangle \langle F^{{\cal BB}}(t_2; {\bf p};\Gamma_4)\rangle} \right]^{1/2},
\end{split}
\end{align}
where the baryonic two-point and three-point correlation functions are respectively defined as:
\allowdisplaybreaks{
\begin{align}
	\begin{split}\label{twopcf}
	&\langle F^{{\cal BB}}(t; {\bf p};\Gamma_4)\rangle=\sum_{\bf x}e^{-i{\bf p}\cdot {\bf x}}\Gamma_4^{\alpha\alpha^\prime} \times \langle \text{vac} | T [\eta_{\cal B}^\alpha(x) \bar{\eta}_{{\cal B}}^{\alpha^\prime}(0)] | \text{vac}\rangle,
	\end{split}\\
	\begin{split}\label{thrpcf}
	&\langle F^{{\cal B V_\mu B^\prime}}(t_2,t_1; {\bf p}^\prime, {\bf p};\mathbf{\Gamma})\rangle=-i\sum_{{\bf x_2},{\bf x_1}} e^{-i{\bf p}\cdot {\bf x_2}} e^{i{\bf q}\cdot {\bf x_1}} \Gamma^{\alpha\alpha^\prime} \langle \text{vac} | T [\eta_{\cal B}^\alpha(x_2) V_\mu(x_1) \bar{\eta}_{{\cal B}^\prime}^{\alpha^\prime}(0)] | \text{vac}\rangle,
	\end{split}
\end{align}
}%
with $\Gamma_i=\gamma_i\gamma_5\Gamma_4$ and $\Gamma_4\equiv (1+\gamma_4)/2$. 

The baryon interpolating fields are chosen, similarly to that of the octet baryons, as
\begin{align}
		\eta_{\xcc}(x)=\epsilon^{ijk}[c^{T i}(x) C \gamma_5 \ell^j(x)]c^k(x),\\
		\eta_{\sgc}(x)=\epsilon^{ijk}[\ell^{T i}(x) C \gamma_5 c^j(x)]\ell^k(x),\\
		\eta_{\occ}(x)=\epsilon^{ijk}[c^{T i}(x) C \gamma_5 s^j(x)]c^k(x),\\
		\eta_{\omc}(x)=\epsilon^{ijk}[s^{T i}(x) C \gamma_5 c^j(x)]s^k(x),
\end{align}
where $\ell=u$ for the doubly charged $\Xi_{cc}^{++}(ccu)$/$\Sigma_{c}^{++}(cuu)$ and $\ell=d$ for the singly charged $\Xi_{cc}^{+}(ccd)$/$\Sigma_{c}^{+}(cdd)$ baryons. Here $i$, $j$, $k$ denote the color indices and $C=\gamma_4\gamma_2$. $t_1$ is the time when the external electromagnetic field interacts with a quark and $t_2$ is the time when the final baryon state is annihilated. When $t_2-t_1$ and $t_1\gg a$, the ratio in Eq.~(\ref{ratio}) reduces to the desired form
\begin{equation}\label{desratio}
	R(t_2,t_1;{\bf p^\prime},{\bf p};\Gamma;\mu)\xrightarrow[t_2-t_1\gg a]{t_1\gg a} \Pi({\bf p^\prime},{\bf p};\Gamma;\mu).
\end{equation}
We extract the form factors $G_{E,\cal{B}}(q^2)$ and $G_{M,\cal{B}}(q^2)$ by choosing appropriate combinations of Lorentz direction $\mu$ and projection matrices $\Gamma$:
\begin{align}
	\Pi({\bf 0},{\bf -q};\Gamma_4;\mu=4)&=\left[\frac{(E_{\cal B}+m_{\cal B})}{2E_{\cal B} }\right]^{1/2} G_{E,\cal{B}}(q^2),\\
	\Pi({\bf 0},{\bf -q};\Gamma_j;\mu=i)&=\left[\frac{1}{2E_{\cal B} (E_{\cal B}+m_{\cal B})}\right]^{1/2} \epsilon_{ijk}\, q_k\, G_{M,\cal{B}}(q^2).
\end{align}
Here, $G_{E,\cal{B}}(0)$ gives the electric charge of the baryon. Similarly, the magnetic moment can be obtained from the magnetic form factor $G_{M,\cal{B}}$ at zero momentum transfer.
\end{widetext}

\subsection{Lattice setup}
We refer the reader to Refs.~\cite{Can:2012tx, Can:2013zpa} for the details of our lattice setup. We use four sets of configurations with different light quark hopping parameters $\kappa^{u,d}_{sea}= {0.13700,0.13727,0.13754,0.13770}$, which correspond to pion masses of ~ 700, 570, 410 and 300 MeV, respectively. The strange quark mass is fixed to its physical value at $\kappa^s_{sea}= 0.13640$. In order to be consistent with the sea quarks we use the clover action for the u, d and s valence quark propagators and their $\kappa$ values are chosen to be the same $\kappa^q_{sea}=\kappa^q_{val}$. 

We employ a wall method~\cite{Can:2012tx}, which provides a simultaneous study of all the hadrons we are considering. The smeared source and wall sink are separated by 12 lattice units in the temporal direction. We smear the delta function source operator over the three spatial dimensions of the source time slice in a gauge-invariant manner using a Gaussian form. In the case of $u$, $d$ and $s$ quarks, we choose the smearing parameters so as to give a root-mean-square radius of $\langle r_l \rangle \sim 0.5$~fm. As for the charm quark, we adjust the smearing parameters to obtain $\langle r_c \rangle=\langle r_l \rangle/3$.

For the charm quarks, we apply the Fermilab method~\cite{ElKhadra:1996mp} in the form employed by the Fermilab Lattice and MILC Collaborations~\cite{Burch:2009az, Bernard:2010fr}. A similar approach has been recently used to study charmonium, heavy-light meson resonances and their scattering with pion and kaon~\cite{Mohler:2011ke, Mohler:2012na, Mohler:2013rwa}. In this simplest form of the Fermilab method, the clover coefficients $c_E$ and $c_B$ in the action are set to the tadpole-improved value $1/u_0^3$, where $u_0$ is the average link. Following the approach in Ref.~\cite{Mohler:2011ke}, we estimate $u_0$ to be the fourth root of the average plaquette. 

\begin{figure}[t]
	\centering
	\includegraphics[width=0.5\textwidth]{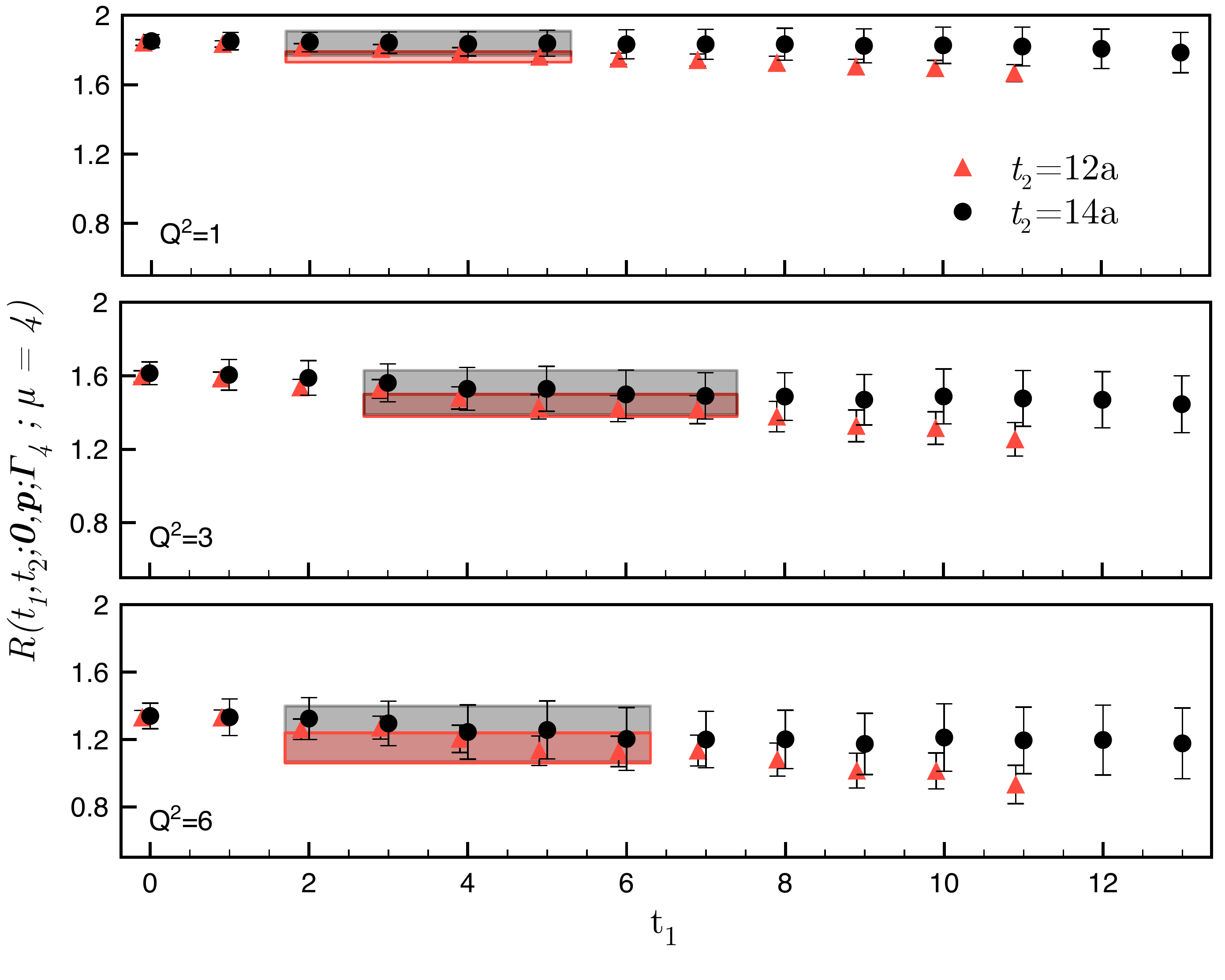}
	\caption{\label{E12_14} The ratio in Eq.~\eqref{ratio} as function of the current insertion time, $t_1$, for the electric form factor of $\Xi_{cc}$ with $t_2=12a$ and $t_2=14a$. We show statistics over 30 configurations for three illustrative momentum-transfer values. The data for $t_2=12a$ are slightly shifted to left for clear viewing.}
\end{figure}	
\begin{figure}[t]
	\centering
	\includegraphics[width=0.5\textwidth]{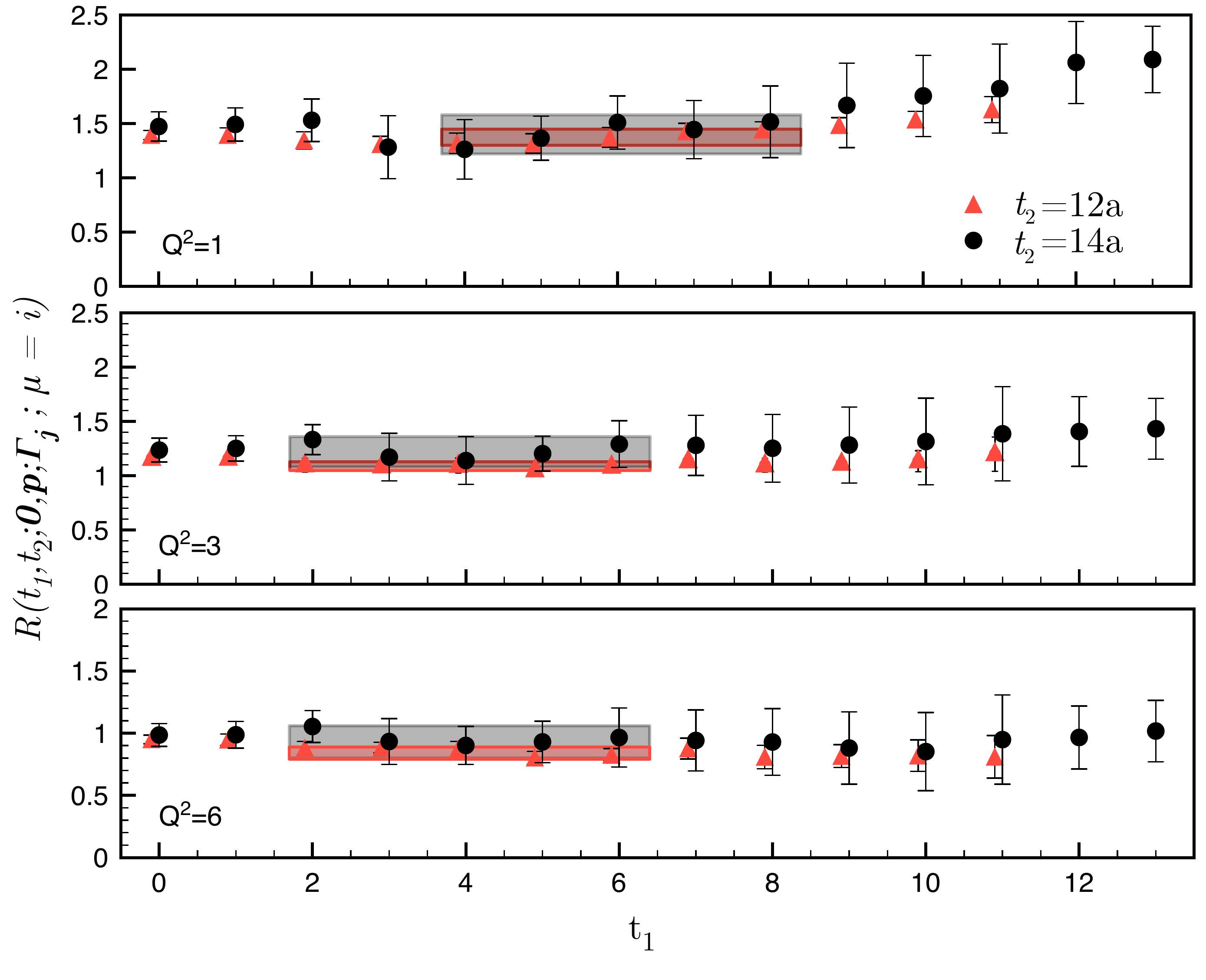}	
	\caption{\label{M12_14} Same as Fig.~\ref{E12_14} but for the magnetic form factor of $\Xi_{cc}$.}
\end{figure}	
\begin{figure}[t]
	\centering
	\includegraphics[width=0.5\textwidth]{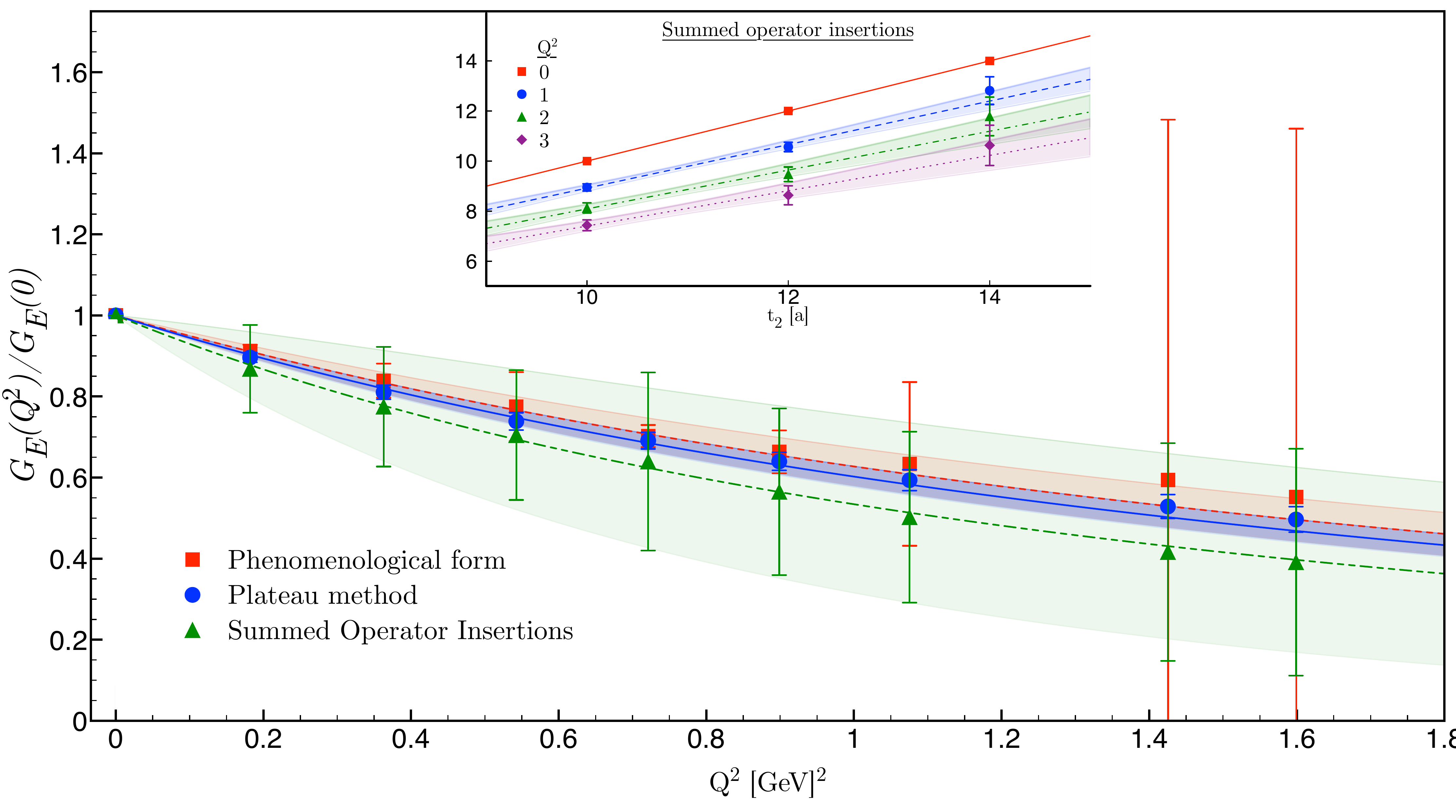}	
	\caption{\label{Esummed} A comparison of the electric form factor of $\Xi_{cc}$ for the heaviest quark mass, as obtained using a simple plateau fit, the phenomenological fit form in Eq.~\ref{excfit} and the summation method. The small panel depicts the summed operator insertions for three time separations and for the first four momentum insertions with their linear fits.}
\end{figure}	

We determine the charm-quark hopping parameter $\kappa_c$ nonperturbatively. To this end, we measure the spin-averaged static masses of charmonium and heavy-light mesons and tune their values accordingly to the experimental results. We perform our simulations with $\kappa_c=0.1246$, which gives the ground-state masses as listed in Table~\ref{res_table1}. Baryon spectrum will be further discussed in Section~\ref{sec3}.

For each $\kappa^{ud}_{sea}$ value, we perform our measurements on 100, 100, 150 and 170 different configurations for the $\Sigma_c$ and 100, 100, 100 and 130 different configurations for the $\Omega_c$ and $\Omega_{cc}$ baryons. In order to increase the statistics for the $\Sigma_c$ and $\Xi_{cc}$ baryons , we have employed multiple source-sink pairs by shifting them 12 lattice units in the temporal direction while one pair have been enough for the others. We insert momentum through the current up to nine units: $(|p_x|,|p_y|,|p_z|)$=(0,0,0), (1,0,0), (1,1,0), (1,1,1), (2,0,0), (2,1,0), (2,1,1), (2,2,0), (2,2,1) and average over equivalent momenta in the case of electric form factor. For the magnetic form factor we average over all equivalent combinations of spin projection, Lorentz component and momentum indices. All statistical errors are estimated by the single-elimination jackknife analysis and the $\chi^2$ p-values and Akaike Information Criterion are used to test the goodness of fits and models. 

For the vector current, we consider both the local,
\begin{equation}
	V_\mu=\overline{q}(x)\gamma_\mu q(x),
\end{equation}
and the point-split lattice current,
\begin{equation}
V_\mu = 1/2[\overline{q}(x+\mu)U^\dagger_\mu(1+\gamma_\mu)q(x) -\overline{q}(x)U_\mu(1-\gamma_\mu)q(x+\mu)],
\end{equation}
which is conserved by the Wilson fermions, therefore does not require any renormalisation. Both results are in good agreement, thus we report only the point-split one.

In our simulations, the source-sink time separation is fixed to ~1.09 fm ($t_2=12a$). Statistics limit the upper value of $t_2$; as we increase the separation the statistical errors grow rapidly. Therefore, we must choose the smallest possible separation value ensuring that the excited-state contaminations are avoided. As for the nucleon axial and electromagnetic form factors, a separation of ~1 fm has been found to be sufficient~\cite{Alexandrou:2010hf, Alexandrou:2011db}. A similar conclusion has been made for the $\Omega^-$ electromagnetic form factors~\cite{Alexandrou:2010jv}. To check that a separation of $t_2=12a$ is sufficient for the charmed baryons, we compared our results with those obtained using a separation of $t_2=14a$. As an illustration, in Figs.~\ref{E12_14} and \ref{M12_14} we show the ratio in Eq.~\eqref{ratio} as function of the current insertion time, $t_1$, for the electric and magnetic form factors of $\Xi_{cc}$ with $t_2=12a$ and $t_2=14a$. As can be seen, the plateau values obtained from the two time separations are consistent with each other, implying that the shorter source-sink time separation is sufficient. As compared to $t_2=12a$, the error bars for $t_2=14a$ are at least twice as large. Other baryons we study exhibit a similar behavior, therefore we use the shorter separation \emph{i.e.} $t_2=12a$ in all of our analysis.

To further ensure that the ground baryon state is isolated from the excited-state contaminations we performed a secondary analysis and fitted the ratio in Eq.~\eqref{ratio} to a phenomenological form 
\begin{equation}\label{excfit}
	R(t_2,t_1)=G_{E,M}+b_1\,e^{-\Delta t_1}+b_2\,e^{-\Delta (t_2-t_1)},
\end{equation}
where $\Delta$ is the energy gap between the ground and the excited state. In the case of nucleon form factors using the sequential-source inversion method, this approach has proved to be useful in a more systematic analysis accounting for the excited-state contaminations (see \emph{e.g.} Ref.~\cite{Bhattacharya:2013ehc} for a rigorous test). One obstacle we have in the case of charmed baryons is that the energy gaps are unknown. Hence we take $\Delta$ as a free parameter together with $b_1$ and $b_2$, yielding a larger uncertainty for $G_{E,M}$. One other caveat is that the source and the sink we utilize are asymmetric in smearing, which implies $b_1\neq b_2$. 

\begin{sidewaysfigure*}[p]
	\centering
	\vspace{8cm}
	\includegraphics[width=1\textwidth]{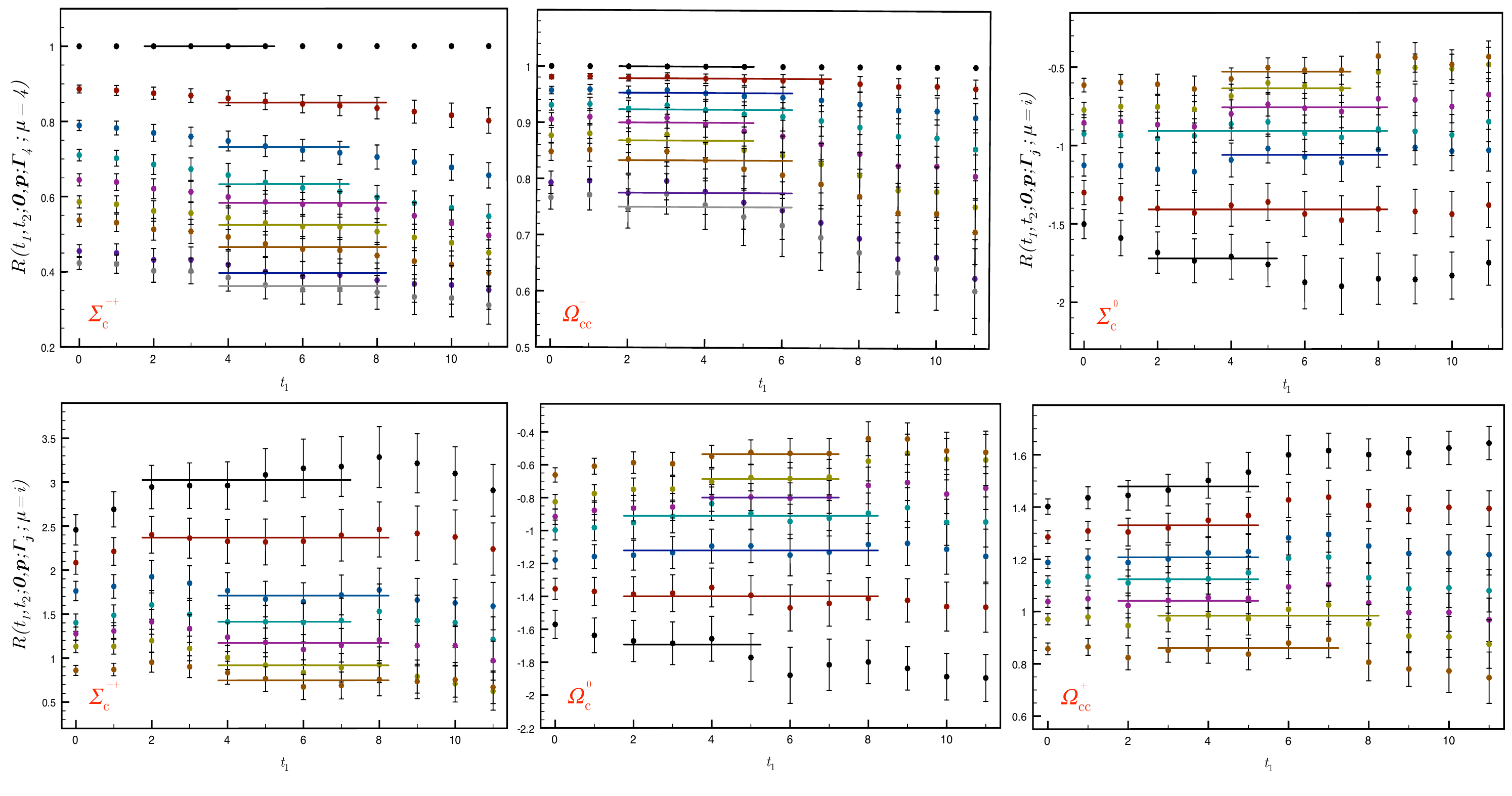}
	\caption{\label{el_plat} The ratio in Eq.~\eqref{ratio} as function of the current insertion time, $t_1$, for the electric form factors of $\sgc^{++}$ and $\occ^{+}$ as normalized with their electric charges, and for the magnetic form factors of $\sgc^0$, $\sgc^{++}$, $\omc^0$ and $\occ^{+}$. We show the data for $\kappa_\text{val}=0.13700$ only and for first nine four\--momentum insertions. The horizontal lines denote the plateau regions as determined by using a p-value criterion (see text).}
\end{sidewaysfigure*}	

\begin{sidewaysfigure*}[p]
	\centering
	\vspace{8cm}	
	\hspace{-3cm}
	\includegraphics[width=1.1\textwidth]{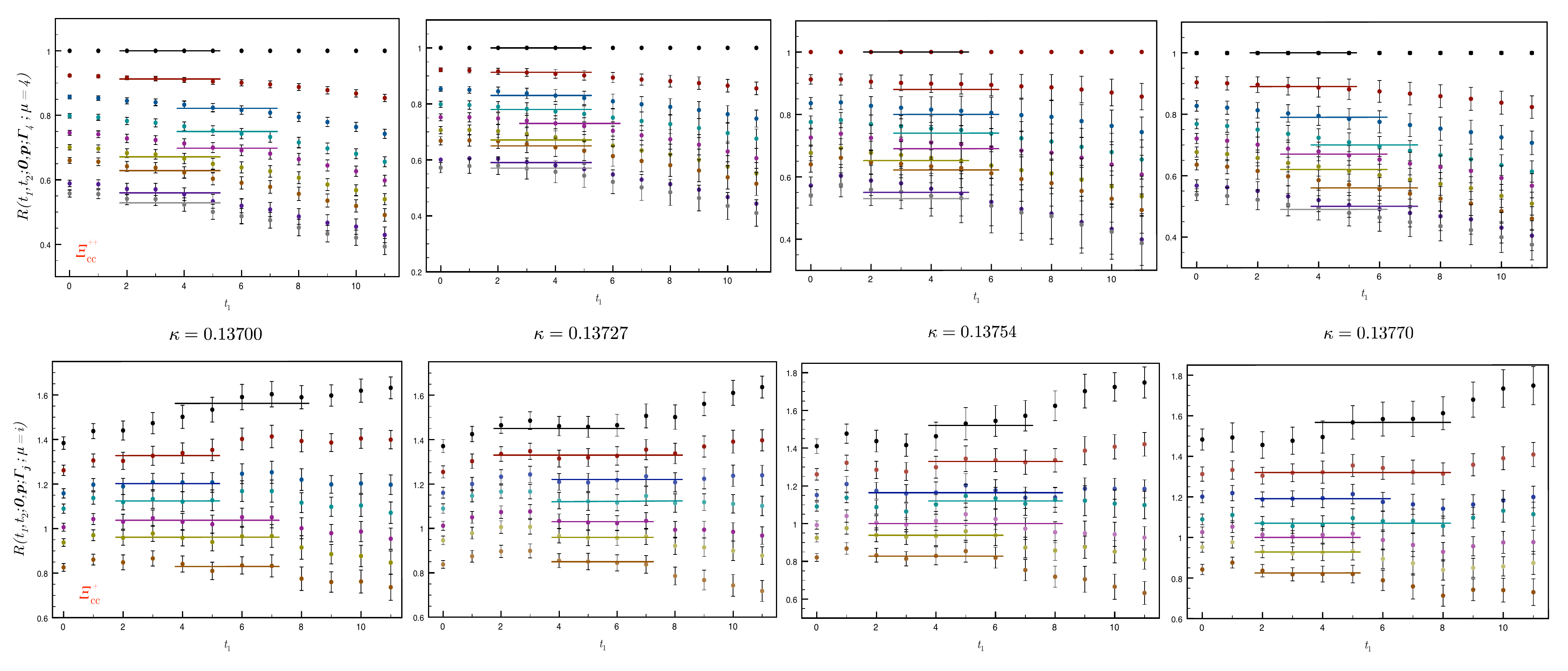}
	\caption{\label{xi_plat} Same as Fig.~\ref{el_plat} but for the electric form factors of $\Xi_{cc}^{++}$ and for the magnetic form factors of $\Xi_{cc}^+$. We show the data $\kappa_\text{val}$ values we consider.}
\end{sidewaysfigure*}	



In Fig.~\ref{Esummed} we compare the electric form factor of $\Xi_{cc}$ as obtained using a simple plateau fit and the phenomenological fit form in Eq.~\eqref{excfit}, for three illustrative momentum transfers and for the heaviest quark mass. It can be seen that the two fit forms give completely consistent results, the error bars being twice as large for the phenomenological fit form. In Table~\ref{exp_fit_res_table} we give the parameter values of $R(t_2,t_1)$ in the case of electric form factors of $\Xi_{cc}$ for all  momentum transfers. The statistical error in $\Delta$ values is quite large as expected. Note that we do not intend to interpret $\Delta$ as the physical energy gap at this stage. On the other hand we have not been able to obtain a good fit to the phenomenological form for the magnetic form factors. For all the momentum transfers, the statistical errors in the parameters of the fit to $R(t_2,t_1)$ are too large to allow a precise determination of $G_M$. Therefore, we use solely a plateau fit in extracting the ground state matrix elements of electric and magnetic form factors.

\begin{table}[ht]
	\caption{ The parameter values of $R(t_2,t_1)$ in the case of electric form factors of $\Xi_{cc}$ for all momentum transfers and for the heaviest quark mass.  
}
\begin{center}
	{
	\setlength{\extrarowheight}{5pt}
\begin{tabular*}{0.5\textwidth}{@{\extracolsep{\fill}}c|cccc}
			\hline\hline 
			$Q^2$  & $G_E(Q^2)$ & $\Delta$ &  $b_1$ &   $b_2$    \\
			\hline \hline
			[$\frac{2\pi}{a\, N_s}$] & & [a] &   \\
			1 & 0.910(18) & 0.220(100)  & 0.026(18) & -0.087(18) \\
			2 & 0.839(42) & 0.186(90)   & 0.045(34) & -0.143(35) \\
			3 & 0.775(85) & 0.169(110)  & 0.064(56) & -0.174(71) \\
			4 & 0.703(27) & 0.285(108)  & 0.082(33) & -0.154(22) \\
			5 & 0.664(53) & 0.212(122)  & 0.079(47) & -0.171(41) \\
			6 & 0.634(202) & 0.174(160)  & 0.078(113) & -0.197(169) \\
			8 & 0.594(889) & 0.145(251)  & 0.062(418) & -0.210(672) \\
			9 & 0.552(908) & 0.160(356)  & 0.066(466) & -0.205(700)  \\
			\hline \hline			
\end{tabular*}
	\label{exp_fit_res_table}
	}
\end{center}
\end{table}

One strategy that can be used to remove the excited-state contaminations is to vary the source-sink separation and extract the ground state matrix elements by using \emph{summed operator insertions}~\cite{Maiani:1987by}. This method has the advantage of computing matrix elements with reduced excited-state contaminations, however it is computationally more demanding as multiple source-sink separations need to be used. To further check that we avoid excited-state contaminations, we have computed the ratio in Eq.~\eqref{ratio} for three source-sink separations, namely for $t_2=10a$, $t_2=12a$ and $t_2=14a$ (for the heaviest quark mass and using 30 configurations), and summing the operator insertions we extracted the form factors. Fig.~\ref{Esummed} depicts also a comparison of the summation method with the other methods. In general, the statistical errors for the summation method are larger, however, the data are consistent within the errors. We intend to study the summation method further with increased statistics in a future work.

\section{Results and Discussion}\label{sec3}

\begin{table*}[ht]
	\caption{ The charmed meson and baryon masses at each quark mass we consider, with their chiral fits to linear and quadratic forms. We also give the experimental values and PACS-CS results for comparison.  
}
\begin{center}
	{
	\setlength{\extrarowheight}{5pt}
\begin{tabular*}{1.0\textwidth}{@{\extracolsep{\fill}}c|cccccc}
			\hline\hline 
			$\kappa^{u,d}_{val}$  &  $\,m_{\eta_c}$ & $\,m_{J/\Psi}$ &  $\,m_{D}$ &   $\,m_{D^\ast}$ &  $\,m_{D_s}$ &   $\,m_{D_s^\ast}$   \\
			\hline \hline
			        & [GeV]     & [GeV]     & [GeV]         & [GeV]   & [GeV]  & [GeV]    \\
			0.13700 & 3.019(3) & 3.116(5)  & 2.027(5)  & 2.180(10)  & 2.075(5) & 2.220(9)\\
			0.13727 & 3.006(3) & 3.097(4)  & 1.982(5)  & 2.112(12)  & 2.052(4) & 2.179(8)\\
			0.13754 & 2.992(3) & 3.079(4)  & 1.934(8)  & 2.077(16) &  2.033(5) & 2.155(8)\\
			0.13770  & 2.984(2) & 3.071(3) & 1.915(9)  & 2.045(16)  & 2.028(4) & 2.156(7)\\
			\hline
			Lin. Fit  & 2.979(2)& 3.063(3)  & 1.895(6) &  2.021(13) & 2.018(4) & 2.138(7)\\
			Quad. Fit  & 2.977(4)& 3.064(5) & 1.893(9) &  2.035(22) & 2.022(7) & 2.156(13)\\
			\hline\hline
			Exp.      & 2.980 & 3.097 & 1.865 & 2.007 & 1.968 &2.112\\
			PACS-CS~\cite{Namekawa:2011wt}    & 2.986(1)(13) & 3.094(1)(14) & 1.871(10)(8) &  1.994(11)(9) & 1.958(2)(9) & 2.095(3)(10)  \\
		\hline\hline 
		$\kappa^{u,d}_{val}$  &  $\,m_{\Sigma_{c}}$ & $\,m_{\Omega_{c}}$ &  $\,m_{\Xi_{cc}}$ &   $\,m_{\Omega_{cc}}$  &   \\
		\hline \hline
		        & [GeV]     & [GeV]     & [GeV]         & [GeV]    &       \\
		0.13700 & 2.841(18) & 2.959(24)  & 3.810(12)  & 3.861(17)  &  \\
		0.13727 & 2.753(19) & 2.834(19)  & 3.740(13)  & 3.806(12)  &  \\
		0.13754 & 2.647(19) & 2.815(26)  & 3.708(16)   & 3.788(16) &   \\
		0.13770  & 2.584(28) & 2.781(26) & 3.689(18)  & 3.781(28)  &  \\
		\hline
		Lin. Fit  & 2.553(18)& 2.740(24)  & 3.660(14) &  3.755(18) & \\
		Quad. Fit  & 2.525(38)& 2.740(67) & 3.687(24) &  3.791(36) &  \\
		\hline\hline
		Exp.      & 2.455     & 2.695      & 3.519 &  -   & \\
		PACS-CS~\cite{Namekawa:2013vu}    & 2.467(39)(11) & 2.673(5)(12) & 3.603(15)(16) &  3.704(5)(16)  & \\
		\hline\hline 

\end{tabular*}
	\label{res_table1}
	}
\end{center}
\end{table*}

\subsection{Baryon masses}
We need the baryon-mass values at each quark mass in order to extract the electromagnetic form factors. We evaluate the baryon masses using the two-point correlator in Eq.~\eqref{twopcf} and give our results in Table~\ref{res_table1}. While we do not use in our analysis, we perform a chiral extrapolation with functions linear and quadratic in $m_\pi^2$ (see the discussion below for chiral extrapolation) and include the results in Table~\ref{res_table1}. 

It is interesting to compare our results for the baryon masses with those obtained by PACS-CS from the same lattices. It must, however, be noted that PACS-CS uses a somewhat different relativistic heavy-quark action for the $c$-quark to keep the $\mathcal{O}(m_Q a)$ errors under control and extracts the masses at the physical point without any chiral extrapolation. Such differences between two analyses need to be taken into account as a source of systematic error. Yet, a mass determination, of course, requires a more systematic chiral fit than linear or quadratic forms as we perform here. Nevertheless, we think such a comparison is useful to see the effect of the discretization errors in our analysis. 

The chiral extrapolations in linear and quadratic forms are consistent with each other within their error bars. For all baryons, we either see an agreement within error bars or only a few percent discrepancy in baryon masses between PACS-CS and our results. This suggests that the discretization errors are relatively small. It is reasonable to expect this effect to be much smaller in the case of form factors which are less sensitive to the charm-quark mass. We actually confirmed this by varying the charm-quark mass so as to change the $\Xi_{cc}$ mass by approximately 100 MeV, which resulted in only a minor change in the charge radii and the magnetic moments.

\begin{figure*}[pt]
	\centering
	\includegraphics[width=1\textwidth]{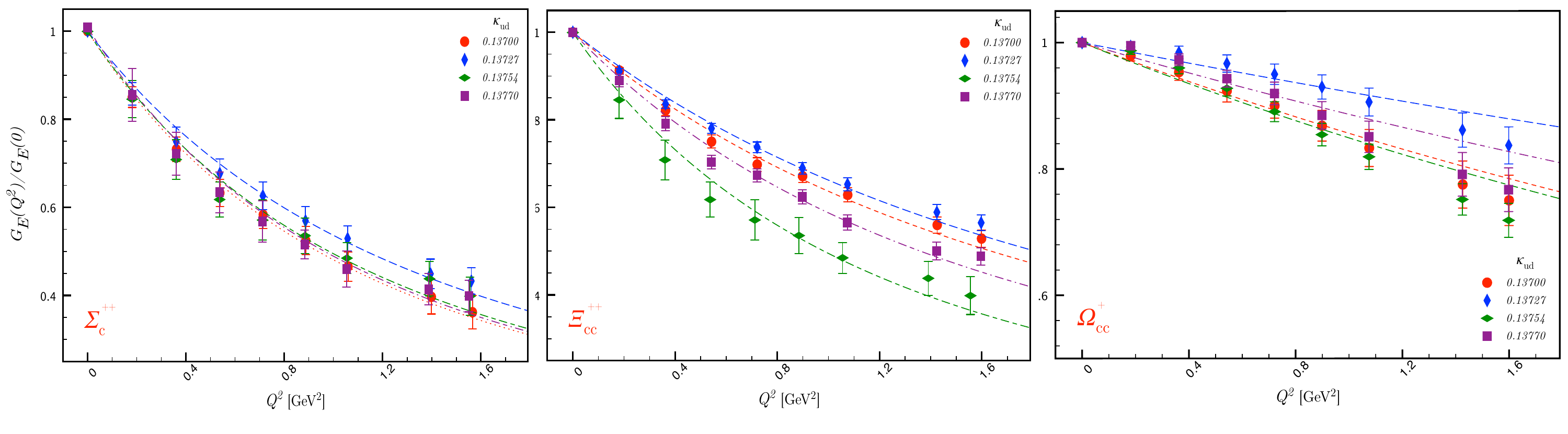}
	\caption{\label{e_dipol} The electric form factors of $\sgc^{++}$,$\Xi_{cc}^{++}$ and $\occ^{+}$ as normalized with their electric charges as functions of $Q^2$, for all the quark masses we consider. The dots mark the lattice data and the curves show the best fit to the dipole form in Eq.~\eqref{dipole}.}
\end{figure*}	

\begin{figure*}[pt]
	\centering
	\includegraphics[width=1\textwidth]{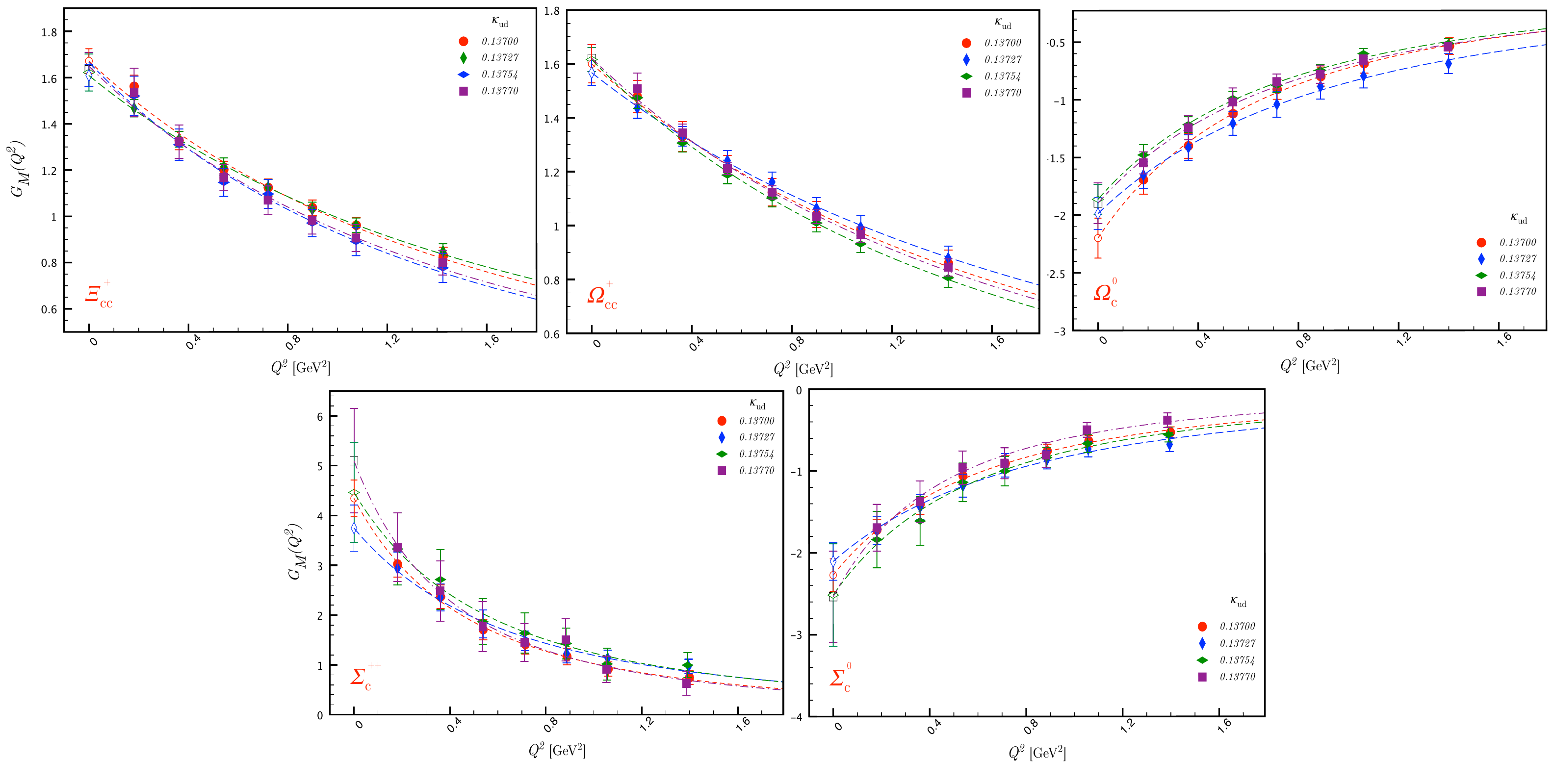}
	\caption{\label{m_dipol} Same as Fig.~\ref{e_dipol} but for the magnetic form factors of $\Xi_{cc}^{+}$, $\occ^{+}$, $\omc^0$, $\sgc^0$ and $\sgc^{++}$ as functions of $Q^2$, for all the quark masses we consider.}
\end{figure*}	

Fig.~\ref{el_plat} shows the ratio in Eq.~\eqref{ratio} as function of the current insertion time, $t_1$, for the electric form factors of $\sgc^{++}$ and $\occ^{+}$, as normalized with their electric charges, and for the magnetic form factors of $\sgc^0$, $\sgc^{++}$, $\omc^0$ and $\occ^{+}$. We present the data solely for $\kappa_{val}=0.13700$ and for the first nine four\--momentum insertions. In order to illustrate how plateau regions change as we approach the chiral limit, we give in Fig.~\ref{xi_plat} the electric form factors of $\Xi_{cc}^{++}$ and the magnetic form factors of $\Xi_{cc}^+$ for all the quark masses we consider.

In determining a plateau region, we consider the p-value as a criterion~\cite{PhysRevD.86.010001, Can:2013zpa}. In each case, we search for plateau regions of minimum three time slices between the source and the sink, and we choose the one that has the highest p-value. The fairly significant time dependence at late times can be explained by the weak coupling to the ground state and resulting excited-state contamination. Therefore the regions closer to the smeared source are preferred as they are expected to couple to the ground state with higher strength as compared to the wall sink.

\begin{figure}[th]
	\centering
	\includegraphics[width=0.4\textwidth]{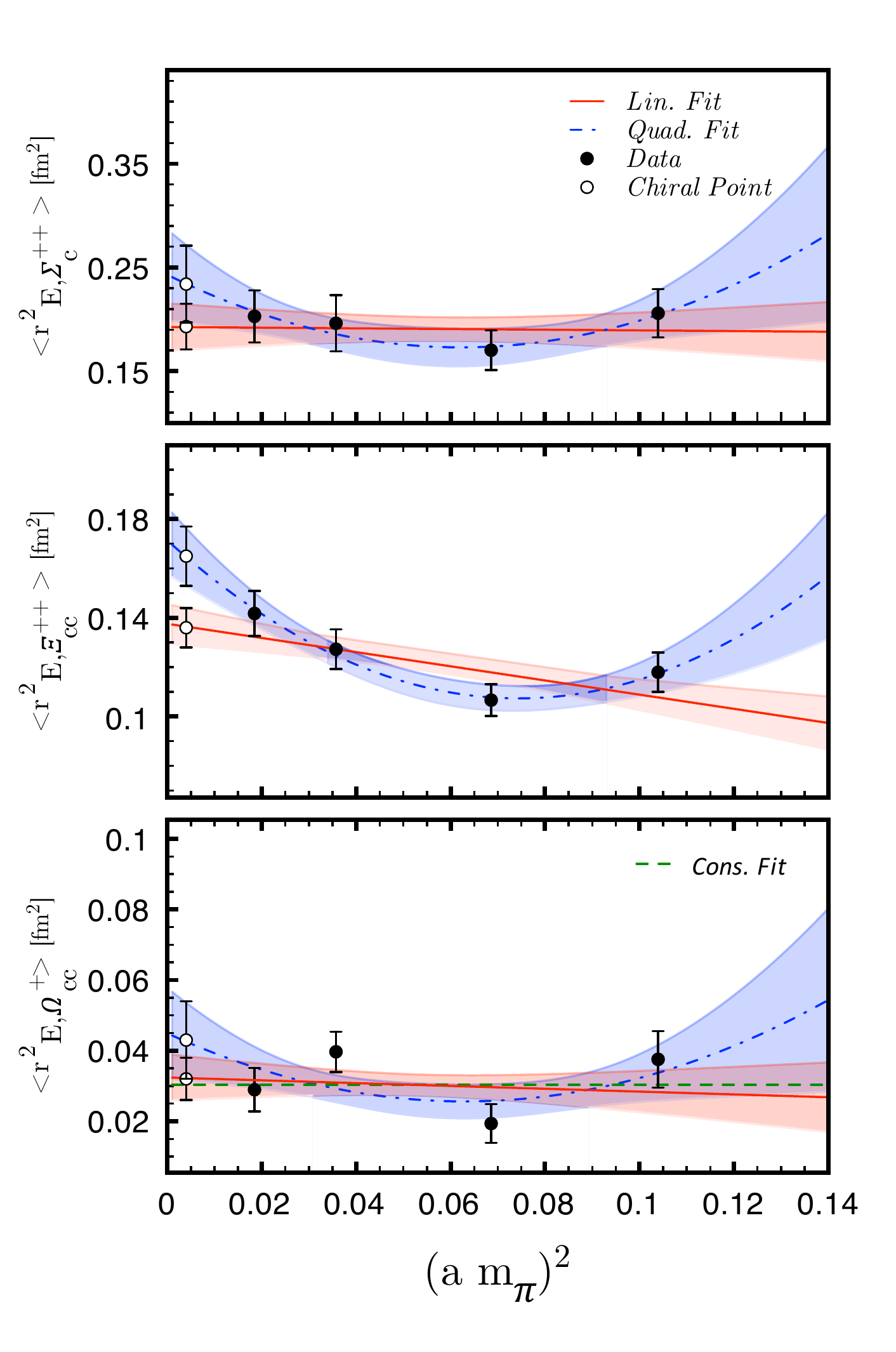}
	\caption{\label{chiral_fit_el} The chiral extrapolations for electric charge radii of $\sgc^{++}$, $\Xi_{cc}^{++}$ and $\occ^{+}$ in $(am_\pi)^2$. We show the fits to constant, linear and quadratic forms. The shaded regions are the maximally allowed error regions, which give the best fit to data.}
\end{figure}	

\begin{figure}[th]
	\centering
	\includegraphics[width=0.4\textwidth]{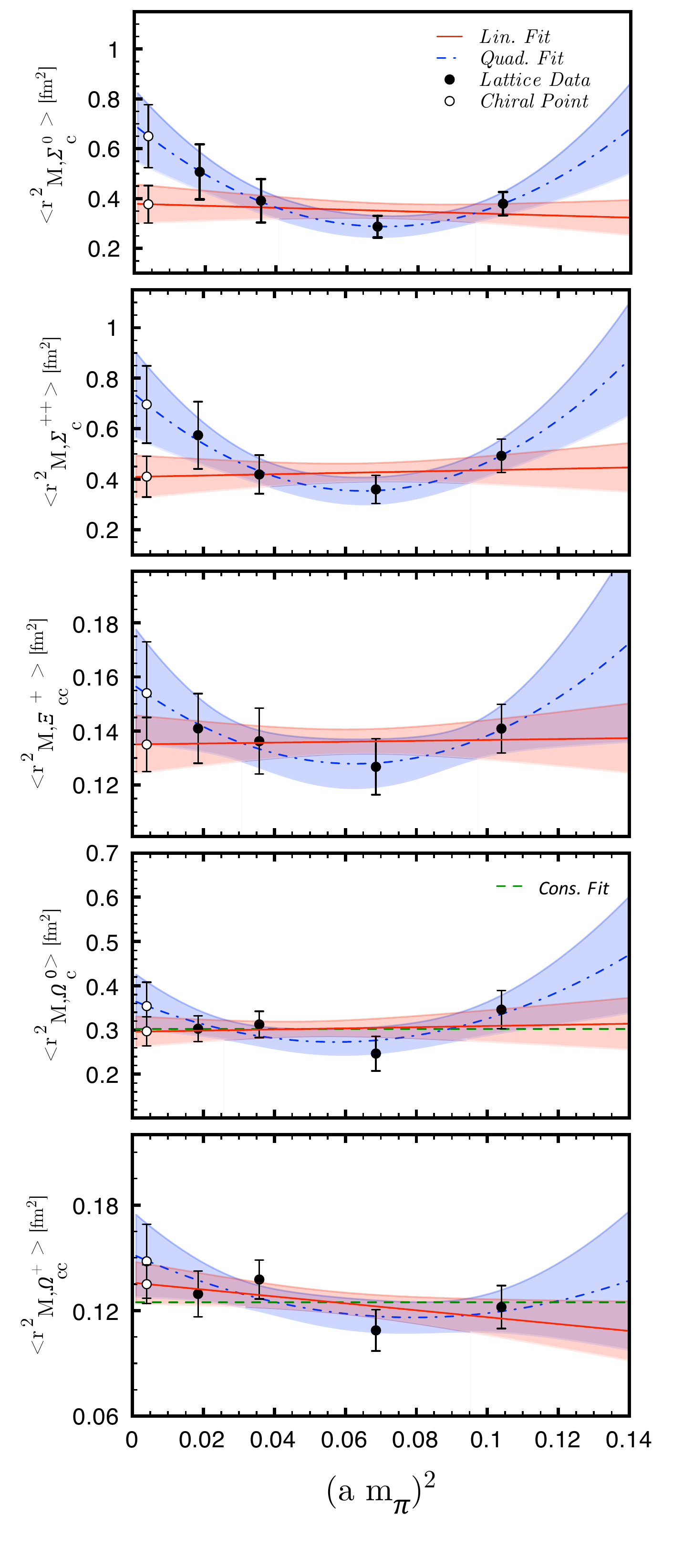}
	\caption{\label{chiral_fit_mag} The chiral extrapolations for magnetic charge radii of $\sgc^{0,++}$, $\Xi_{cc}^{+}$, $\omc^0$ and $\occ^{+}$. We show the fits to constant, linear and quadratic forms.}
\end{figure}	

\begin{figure}[th]
	\centering
	\includegraphics[width=0.4\textwidth]{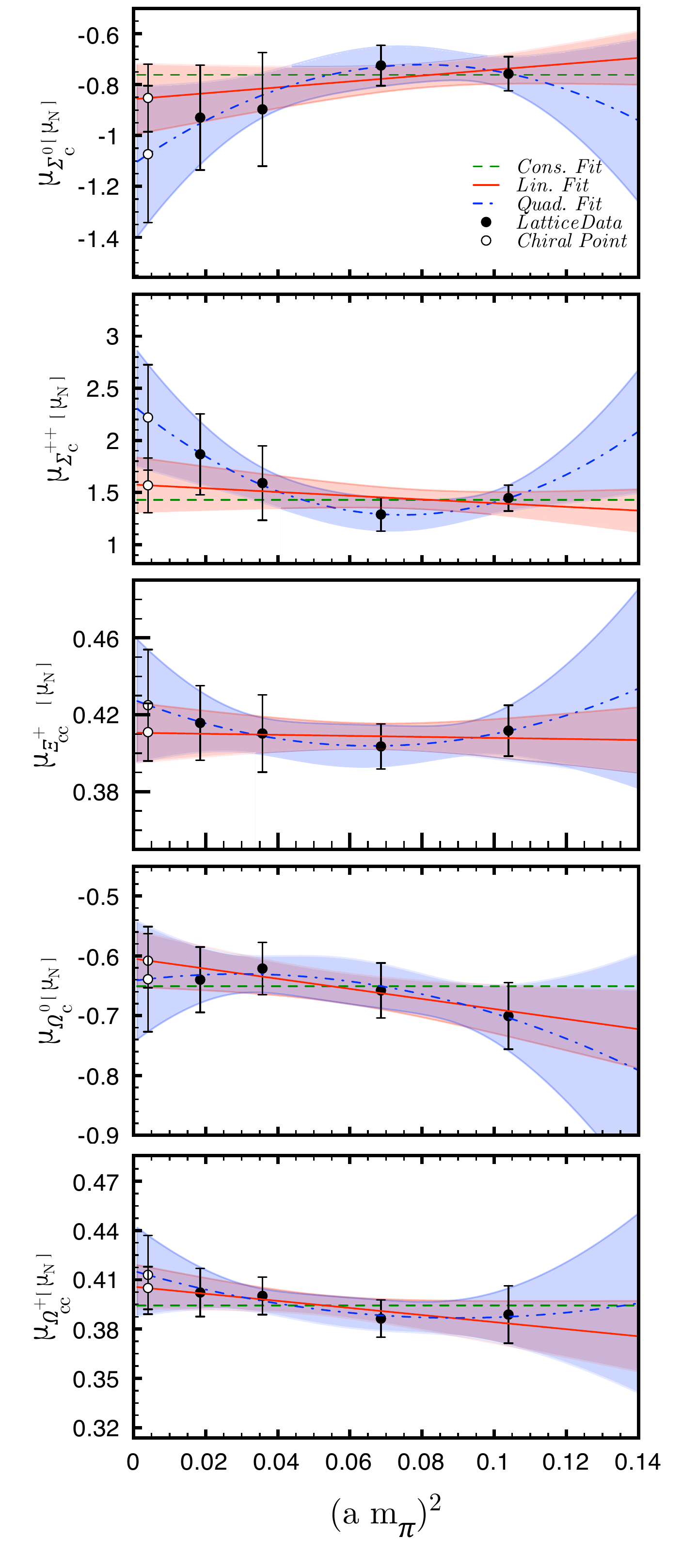}
	\caption{\label{chiral_fit_mom} The chiral extrapolations for magnetic moment of $\sgc^{0,++}$, $\Xi_{cc}^{+}$, $\omc^0$ and $\occ^{+}$. We show the fits to constant, linear and quadratic forms.}
\end{figure}	

\subsection{Lattice evaluation of the data}
In order to evaluate the magnetic moments, we need to extrapolate the magnetic form factor $G_M$ to $-q^2 \equiv Q^2=0$, while the electric charge, which is defined as $G_E(0)$, can be directly computed. We use the following dipole form to describe the $Q^2$ dependence of the baryon form factors:
\begin{equation}\label{dipole}
	G_{E,M}(Q^2)=\frac{G_{E,M}(0)}{(1+Q^2/\Lambda_{E,M}^2)^2}.
\end{equation}
It is well known that this dipole approximation gives a good description of experimental electric form-factor data of the proton. Note that the electric charges of the baryons, $G_E(0)$, are obtained in our simulations to a very good accuracy. 

Fig.~\ref{e_dipol} and~\ref{m_dipol} display the electric form factors of $\sgc^{++}$, $\Xi_{cc}^{++}$ and $\occ^{+}$, as normalized with their electric charges, and the magnetic form factors of $\Xi_{cc}^{+}$, $\occ^{+}$, $\omc^0$, $\sgc^0$ and $\sgc^{++}$ as functions of $Q^2$. We show the lattice data and the fitted dipole forms for all the quark masses we consider. As can be seen from the figures, the dipole form describes the lattice data quite successfully with high-quality fits.

We can extract the electromagnetic charge radii of the baryons from the slope of the form factor at $Q^2=0$,
\begin{equation}
	\langle r_{E,M}^2 \rangle=-\frac{6}{G_\text{E,M}(0)} \frac{d}{dQ^2}G_{E,M}(Q^2)\bigg|_{Q^2=0}.
\end{equation}
To evaluate the charge radii with the above formula, we will use the dipole form in Eq.~\eqref{dipole}, which yields
\begin{equation}\label{emfitform}
	\langle r_{E,M}^2 \rangle=\frac{12}{\Lambda_{E,M}^2}.
\end{equation}
Then the charge radii can be directly calculated using the values of dipole masses as obtained from our simulations.


The magnetic moment is defined as $\mu_{\cal B}=G_{M}(0) e/(2m_{\cal B})$ in natural units. We obtain $G_{M}(0)$ by extrapolating the lattice data to $Q^2=0$ via the dipole form in Eq.~\eqref{dipole} as explained above. We evaluate the magnetic moments in nuclear magnetons using the relation
\begin{equation}
	\mu_{\cal B}=G_{M}(0)\left(\frac{e}{2m_{\cal B}}\right)=G_{M}(0)\left(\frac{m_N}{m_{\cal B}}\right)\mu_N,
\end{equation}
where $m_N$ is the physical nucleon mass and $m_{\cal B}$ is the baryon mass as obtained on the lattice.

Our numerical results for the form factors are given in Tables~\ref{res_table2} and \ref{res_table3}, in Appendix. We give  the electric and magnetic charge radii in fm$^2$, the values of magnetic form factors at $Q^2=0$ ($G_{M,{\cal B}}(0)$) and the magnetic moments ($\mu_{\cal B}$) in nuclear magnetons at each quark mass we consider. These numerical values are illustrated in Figs.~\ref{chiral_fit_el},~\ref{chiral_fit_mag},~\ref{chiral_fit_mom} with their chiral extrapolations for the electric radii, magnetic charge radii and the magnetic moments of the baryons, respectively. To obtain the values of the observables at the chiral point, we perform fits that are constant, linear and quadratic in $m_{\pi}^2$:
\begin{align}
	&f_\text{con}= c_1,\\
	&f_\text{lin}=a_1\,m_\pi^2+b_1,\\
	&f_\text{quad}=a_2\,m_\pi^4+b_2\,m_\pi^2+c_2,
\end{align}
where $a_{1,2},b_{1,2},c_{1,2}$ are the fit parameters. 

In order to evaluate the quality of the fits, we find their $\chi^2$ per degree of freedom value and the p-values. The chiral extrapolations with linear and quadratic forms deviate from each other with their one to two standard deviations in some cases, in particular for $\sgc$. A closer inspection with the $\chi^2$ per degree of freedom and the p-values taken into account reveals that the quadratic form is favored in the case of charge radii and the linear form is favored in the case of magnetic moments.

In the case of charmed-strange baryons $\Omega_c$ and $\Omega_{cc}$, the pion-mass dependence is solely due to sea-quark effects. As can be seen in the lowest panels of Fig.~\ref{chiral_fit_el} and Fig.~\ref{chiral_fit_mag}, the dependence of charge radii for these baryons fluctuates as we approach the chiral limit, in contrary to the naive expectation. This fluctuation may also be due to uncontrolled systematic errors. An intuitive model is to fit these data to a constant or a linear form, since a more complex form is not known for sea-quark dependence. Unfortunately, the fluctuating data results in a poor fit to a linear or quadratic form in the case of $\Omega_c$ and $\Omega_{cc}$ charge radii. Note that the data in other cases can be nicely fit to linear or quadratic forms.

In assessing the best fit function to data, we also account for the consistency between the properties of the baryons as extrapolated to the quark-mass point $m_\pi^2=m_{\eta_{ss}}^2$. Unfortunately we do not have the value of $m_{\eta_{ss}}$ at the SU(3) symmetric point. However, we can make an estimation using the value $m_{\eta_{ss}}=0.39947$, which was extracted by PACS-CS on a lattice with $\kappa_{ud}=\kappa_\text{sea}=0.13700$ and $\kappa_s=0.13640$. The charge radii and the magnetic moments of $\xcc^{+}$ and $\occ^{+}$, as well as those of $\sgc^{0}$ and $\omc^{0}$, are expected coincide at this point. The properties of the $\sgc^{++}$ baryon as extrapolated to this region can be compared with those of an unphysical baryon similar to $\omc^{++}$ but the $s$ quarks are assigned with electric charge 2/3 \textemdash a state that can be easily created on our setup with trivial replacements.

\begin{table*}[ht]
	\caption{ Comparison of our results with various other models. All values are given in nuclear magnetons [$\mu_N$].}
\begin{center}
	{
	\setlength{\extrarowheight}{5pt}
\begin{tabular*}{1.0\textwidth}{@{\extracolsep{\fill}}ccccccccccccc}
		\hline\hline 
				 & \multicolumn{2}{c}{Our result}  & \cite{JuliaDiaz:2004vh} & \cite{Faessler:2006ft} & \cite{Albertus:2006ya}  & \cite{Bernotas:2012nz} & \cite{Sharma:2010vv} & \cite{Barik:1984tq} & \cite{Kumar:2005ei} & \cite{Patel:2007gx} & \cite{Zhu:1997as}\\
				&  Lin. fit & Quad. fit &&&&&&&& \\
				\hline \hline
				$\mu_{\Sigma^{0}_{c}}$  & -0.875(103) & -1.117(198) & -1.78 & -1.04 & - & -1.043 & -1.60 & -1.391 & -1.17 & -1.015 & -1.6(2)\\
				$\mu_{\Sigma^{++}_{c}}$ & 1.499(202) & 2.027(390)  & 3.07 & 1.76 & - & 1.679 & 2.20 & 2.44 & 2.18 & 2.279 & 2.1(3)\\
				$\mu_{\Xi^{+}_{cc}}$ & 0.411(15) & 0.425(29) & 0.94 & 0.72 & $0.785^{+0.050}_{-0.030}$ & 0.722 & 0.84 & 0.774 & 0.77 & - & - &\\
				$\mu_{\Omega^{0}_{c}}$ & -0.608(45) & -0.639(88) & -0.90 & -0.85 & - & -0.774 & -0.90 & -0.85 & -0.92 & -0.960 & - &\\				
				$\mu_{\Omega^{+}_{cc}}$ & 0.405(13) & 0.413(24) & 0.74 & 0.67 & $0.635^{+0.012}_{-0.015}$ & 0.668 & 0.697 & 0.639 & 0.70 & 0.785 & - &\\
		\hline\hline
	\end{tabular*}
		\label{res_comparison_table}
		}
	\end{center}
	\end{table*}
	
\subsection{Electric properties}

We can compare the electric charge radii of $\occ^+$ and $\xcc^+$~\cite{Can:2013zpa}. They are about the same size, which is much smaller as compared to that of the proton (the experimental value is $\langle r_{E,p}^2 \rangle=$0.770~fm$^2$~\cite{PhysRevD.86.010001}). The $s$ quark in $\occ^+$ seems to have no extra effect on charge radius with respect to the light quark in $\xcc^+$. Of all the four charged baryons ($\sgc^{++}$, $\xcc^{++}$, $\xcc^{+}$ and $\occ^+$) we have studied, $\sgc^{++}$ appears to have the largest charge radius. However, the difference in the charge radii of the doubly charged baryons, $\sgc^{++}$ and $\xcc^{++}$, is quite small. 

To gain a deeper insight to the inner quark dynamics, we examine the contribution of individual quarks to the electromagnetic properties of the baryons. This is done by coupling the electromagnetic field solely to the light quark ($u/d/s$) or the $c$ quark. Table~\ref{res_table_quark} in Appendix displays the radii of light or $c$-quark distributions within the baryons. The light quark distributions are systematically larger than those of the $c$ quark. The heavy $c$ quark core acts to shift the center of mass towards itself reducing the size of the baryon. Note that the $c$-quark distributions do not differ much between the singly and the doubly charmed baryons. Similarly, the $u/d$- and $s$-quark distributions are roughly the same. Therefore, the overall effect is small and $\sgc^{++}$ and $\xcc^{++}$, as well as, $\occ^+$ and $\xcc^+$ have almost the same sizes.

A more pronounced effect can be seen by changing the light quark mass. As the $u/d$ quark in $\sgc$ and $\xcc$ baryons becomes lighter the radius of the light quark increases. This is due to the shift in the center of mass towards the heavy $c$ quark and therefore the light quark has a larger distribution. An unexpected behavior occurs as the $u/d$ quark becomes heavier: Initially the charge radii decrease and as we approach the $s$-quark mass region they start to increase again, which can be described nicely by a quadratic function. We have argued in Ref.~\cite{Can:2013zpa} that such behavior may be related to the modification of the confinement force in the hadrons.

On the other hand, the quark-mass dependence of the $\occ^{+}$ baryon is somewhat unstable and it is not straightforward to make a firm statement. Note that we fix the valence $s$-quark mass and the variation is due to $u/d$ the mass of the quark in the sea only. Fit analyses with linear and quadratic functions reveal slight sea-quark mass dependences, which will be partly due to sea-quark effects.

\subsection{Magnetic properties}

A similar pattern can be seen for the magnetic charge radii of the charmed baryons. A better fit is obtained by a quadratic form for all baryons. $\sgc^{++}$ has the largest magnetic radii. Unfortunately, the errors are too large to make a vigorous comparison with the magnetic radius of the proton. $\sgc^{++}$ and $\sgc^0$ seem to have a similar magnetic radii to that of the proton, which is $\langle r_{M,p}^2\rangle=0.604$~fm$^2$~\cite{PhysRevD.86.010001}. $\occ$ has the smallest magnetic charge radii. An inspection of the $\omc$ and $\occ$ magnetic moments and their dependence on the pion mass, which is due to only sea-quarks, reveals that the moments are almost independent of the sea quark effects.

It is also instructive to study the individual quark sector contributions to the magnetic moments of the baryons. For the singly charmed $\sgc$ and $\omc$ baryons, the light-quark distribution is much larger as compared to that of the heavy quark; their magnetic moments are dominantly determined by the light quark. On the contrary, the individual quark sector distributions to the magnetic moments of the doubly charmed $\xcc$ and $\occ$ are similar in magnitude. It follows that the heavy quark plays an equivalent role with the light quark only when it is doubly represented in the baryons. 

The opposite signs of the light- and heavy-quark magnetic moments indicate that their spins are anti-aligned in the baryon most of the time. The spins of the singly charmed $\sgc$ and $\omc$ baryons are mainly determined by the doubly represented light quarks. Generally speaking, when a quark is doubly represented, the quarks are paired in a spin-1 state with their spins aligned. In the case of the doubly charmed baryons, this leads to a larger heavy-quark contribution to the total spin and magnetic moment. $\sgc^{++}$ has the largest magnetic moment of all and the strange baryons $\omc$ and $\occ$ have somewhat smaller moments. It is interesting to compare these values with the experimental magnetic moment of the proton, which is $\mu_p=2.793~\mu_N$~\cite{PhysRevD.86.010001}.

Table~\ref{res_comparison_table} displays a comparison of our results for the magnetic moments with those from various other models. While the signs of the magnetic moments are correctly determined, there is a large discrepancy among results. For all the baryons, the moments seem to be underestimated with respect to other methods. This is similar to what we have found for the $\xcc^+$ baryon~\cite{Can:2013zpa}.

\section{Summary and Conclusion}~\label{sec4}

We have investigated the electromagnetic properties of the singly charmed $\sgc$, $\omc$ and the doubly charmed $\xcc$, $\occ$ baryons from 2+1-flavor simulations of QCD on a $32^3\times 64$ lattice. We have extracted the electric and magnetic charge radii and the magnetic moments. Our results imply that the charmed baryons are compact with respect to baryons that are composed of only light quarks, \emph{e.g.}, the proton. 

A closer inspection of individual quark sector contributions to the charge radii reveals that the light quark distributions are larger. The heavy quark acts to decrease the size of the baryon to smaller values. The doubly charmed baryons are more compact as compared to singly charmed baryons of the same charge. As the  $u/d$ quark in $\sgc$ and $\xcc$ becomes lighter it is pushed out to a larger distance from the heavy quark and as a result the charge radii increase. As it becomes heavier towards the $s$-quark mass region, the sizes are seen to increase again. This may be due to the modification of confinement forces in the baryon.

The electromagnetic charge radii of $\omc$ and $\occ$ baryons seem to be somewhat dependent on the sea-quark mass. This indicates that the quadratic behavior with respect to changing quark mass is partly due to sea-quark effects. The magnetic moments are seen to be almost independent of such effects.

$\occ$ has the smallest magnetic charge radii among all the baryons. $\sgc^{++}$ and $\sgc^0$ baryons have larger and roughly the same magnetic radii. The magnetic moments are dominantly determined by the light quarks when they are doubly represented. The role of the heavy quark is significantly enhanced in the case of the doubly charmed baryons. The signs of the magnetic moments are correctly reproduced on the lattice. However, in general we see an underestimation of the magnetic moments as compared to what has been found with other theoretical methods.

\acknowledgments
All the numerical calculations in this work were performed on National Center for High Performance Computing of Turkey (Istanbul Technical University) under project number 10462009. The unquenched gauge configurations employed in our analysis were generated by PACS-CS collaboration~\cite{Aoki:2008sm}. We used a modified version of Chroma software system~\cite{Edwards:2004sx}. This work is supported in part by The Scientiﬁc and Technological Research Council of Turkey (TUBITAK) under project number 110T245 and in part by KAKENHI under Contract Nos. 22105503, 24540294 and 22105508.

\appendix*

\section{Numerical results}

To allow the reader to investigate the details of our simulations pertaining the quark mass dependence, chiral extrapolations and individual quark sector contributions to the electromagnetic properties of the charmed baryons, we have tabulated our numerical results of the lattice calculations.

\begin{table*}[ht]
	\caption{The electric and magnetic charge radii in fm$^2$, the values of magnetic form factors at $Q^2=0$ ($G_{M,{\cal B}}(0)$), the magnetic moments in nuclear magnetons, for ${\cal B}\equiv \sgc^{++},\, \sgc^{0}$ and $\Xi^+_{cc}$ at each quark mass we consider.
}
\begin{center}
	{
	\setlength{\extrarowheight}{5pt}
\begin{tabular*}{1.0\textwidth}{@{\extracolsep{\fill}}cccccc}
		\hline\hline 
				$\kappa^{u,d}_{val}$  & $\langle r_{E,\Sigma^{++}_{c}}^2 \rangle$  &   $G_{M,\Sigma^{++}_{c}}$ & $\mu_{\Sigma^{++}_{c}}$ & $\langle r_{M,\Sigma^{++}_{c}}^2 \rangle$     
				\\
				\hline \hline
				        & [fm$^2$]   &  & [$\mu_N$] & [fm$^2$]  \\
				0.13700 & 0.206(23)  & 4.343(371) & 1.447(125) & 0.492(66)  \\
				0.13727 & 0.170(19)  & 3.747(466) & 1.289(161) & 0.360(56) \\
				0.13754 & 0.196(27)  & 4.462(1.003) & 1.591(358) & 0.419(77) \\
				0.13770 & 0.195(34)  & 5.098(1.050) & 1.867(388) & 0.574(133)  \\
				\hline
				Lin. Fit & 0.192(22) & 4.295(700) & 1.569(253) & 0.410(81) \\
				Quad. Fit & 0.234(37) & 6.017 (1.385) & 2.220(505) & 0.696(153)  \\
				\hline\hline
				$\kappa^{u,d}_{val}$  &
				&   $G_{M,\Sigma^{0}_{c}}$ & $\mu_{\Sigma^{0}_{c}}$ & $\langle r_{M,\Sigma^{0}_{c}}^2 \rangle$  &     \\
						\hline \hline
				0.13700 &   & -2.272(199) & -0.757(67) & 0.379(47) & \\
				0.13727 &   & -2.105(230) & -0.724(80) & 0.287(44) & \\
				0.13754 &   & -2.516(627) & -0.897(223) & 0.391(87) &  \\
				0.13770 &   & -2.537(557) & -0.929(206) & 0.507(111) &  \\
				\hline
				Lin. Fit &  & -2.330(368) & -0.852(133) & 0.377(75) & \\
				Quad. Fit &  & -2.891(736) & -1.073(269) & 0.650(126) &  \\
		\hline\hline
		$\kappa^{u,d}_{val}$  &  $\langle r_{E,\Xi^{++}_{cc}}^2 \rangle$  
		&   $G_{M,\Xi^{+}_{cc}}$ & $\mu_{\Xi^{+}_{cc}}$ & $\langle r_{M,\Xi^{+}_{cc}}^2 \rangle$  &     \\
				\hline \hline
		0.13700 & 0.118(8)  & 1.672(53) & 0.412(13) & 0.141(9) & \\
		0.13727 & 0.107(6)  & 1.609(47) & 0.404(12) & 0.127(10) & \\
		0.13754 & 0.127(8)  & 1.622(80) & 0.410(20) & 0.136(12) &  \\
		0.13770 & 0.142(9)  & 1.635(74) & 0.416(19) & 0.141(13) &  \\
		\hline
		Lin. Fit & 0.136(8) & 1.602(58) & 0.411(15) & 0.135(10) & \\
		Quad. Fit& 0.165(12) & 1.670(110)& 0.425(29)& 0.154(19) &  \\
		
\hline\hline
	\end{tabular*}
		\label{res_table2}
    }
	\end{center}
	\end{table*}

\begin{table*}[ht]
		\caption{Same as Table~\ref{res_table1} but for ${\cal B}\equiv \occ,\, \omc$.
	}
		\addtolength{\tabcolsep}{6pt}	
\begin{center}
	{
	\setlength{\extrarowheight}{5pt}
	\begin{tabular*}{1.0\textwidth}{@{\extracolsep{\fill}}ccccccc}
			\hline\hline
				$\kappa^{u,d}_{val}$  & $\langle r_{E,\Omega_{cc}}^2 \rangle$  &   $G_{M,\Omega_{cc}}$ & $\mu_{\Omega_{cc}}$ & $\langle r_{M,\Omega_{cc}}^2 \rangle$      \\
						\hline \hline
				        & [fm$^2$]      &  & [$\mu_N$] & [fm$^2$] &  \\
				0.13700 &  0.038(8) & 1.600(71) & 0.389(18) & 0.122(12) & \\
				0.13727 &  0.019(6) & 1.567(47) & 0.386(11) & 0.109(12) & \\
				0.13754 &  0.040(6)  & 1.616(45) & 0.400(11) & 0.138(11) &  \\
				0.13770 &  0.029(6)  & 1.621(50) & 0.402(15) & 0.130(13) &  \\
				\hline
				Lin. Fit & 0.032(6) & 1.625(47) & 0.405(13) & 0.135(11) & \\
				Quad. Fit & 0.043(11) & 1.662(87) & 0.413(24) & 0.148(21) &  \\
		\hline\hline
				$\kappa^{u,d}_{val}$  &   &   $G_{M,\Omega_{c}}$ & $\mu_{\Omega_{c}}$ & $\langle r_{M,\Omega_{c}}^2 \rangle$      \\
						\hline \hline
				0.13700 &   & -2.199(173) & -0.701(56) & 0.346(43) & \\
				0.13727 &   & -1.987(138) & -0.658(46) & 0.247(39) & \\
				0.13754 &   & -1.863(129) & -0.621(44) & 0.313(30) &  \\
				0.13770 &   & -1.896(176) & -0.640(55) & 0.303(29) &  \\
				\hline
				Lin. Fit &  & -1.773(141) & -0.608(45) & 0.297(33) & \\
				Quad. Fit & & -1.903(276) & -0.639(88) & 0.354(54) &  \\
		\hline\hline
\end{tabular*}
	\label{res_table3}
}
\end{center}
\end{table*}

\begin{table*}[ht]
\caption{Individual quark sector contributions to the electric charge radii, magnetic charge radii and the magnetic moments of the charmed baryons. Note that the numbers are given independently from the electric charge of the individual quarks that compose the baryons.}
\begin{center}
	{
	\setlength{\extrarowheight}{5pt}
\begin{tabular*}{1.0\textwidth}{@{\extracolsep{\fill}}cc|cc|cc|cc|}
	\hline\hline 
			 Baryon & $\kappa_{ud}$ & $\langle r_{E}^2 \rangle_q$  & $\langle r_{E}^2 \rangle_Q$ & $\langle r_{M}^2 \rangle_q$ & $\langle r_{M}^2 \rangle_Q$  & $\mu_q$ & $\mu_Q$ \\
			\hline \hline
			 & &[fm$^2$] & [fm$^2$] & [fm$^2$] & [fm$^2$]  & [$\mu_N$] & [$\mu_N$] \\
			  						& 0.13700  	& 0.289(49) & 0.091(22) & 0.444(55) & 0.067(43) & 2.178(188)& -0.080(16) \\
									& 0.13727  	& 0.273(41) & 0.054(12) & 0.346(52) & 0.063(25) & 2.046(248) & -0.096(16) \\
			$\Sigma_c^{0,++}$		& 0.13754 	& 0.353(65) & 0.042(18) & 0.394(69) & 0.185(152) & 2.427(572) & -0.115(26) \\				
			 						& 0.13770  	& 0.338(60) & 0.057(25) & 0.506(99) & 0.141(165) & 2.581(555) & -0.061(31)\\
			\hline
									& Lin. Fit 	& 0.347(49) & 0.032(18) & 0.403(67) & 0.098(80) & 2.369(362) & -0.099(21)\\
									& Quad. Fit & 0.390(86) & 0.066(32) & 0.604(118) & 0.236(183) & 2.943(732) & -0.059(36)\\										
			\hline \hline
			  						& 0.13700  	& 0.264(29) & 0.071(5) & 0.471(38) & 0.079(7) & -0.474(34)& 0.414(11) \\
									& 0.13727  	& 0.282(25) & 0.056(5) & 0.371(40) & 0.078(8) & -0.416(33) & 0.421(9) \\
			$\Xi_{cc}^{+,++}$		& 0.13754  	& 0.379(38) & 0.063(5) & 0.473(58) & 0.076(8) & -0.434(52) & 0.420(9) \\				
			 						& 0.13770  	& 0.358(38) & 0.080(7) & 0.477(85) & 0.085(8) & -0.471(86) & 0.432(11) \\
			\hline
									& Lin. Fit 	& 0.386(33) & 0.068(5) & 0.426(60) & 0.082(6) & -0.410(51) & 0.430(8) \\
									& Quad. Fit & 0.410(46) & 0.095(9) & 0.612(115) & 0.089(11) & -0.516(117) & 0.433(16)\\														
			\hline \hline
			  						& 0.13700  	& 0.253(35) & 0.074(20) & 0.424(53) & 0.090(37) & 2.080(155) & -0.071(17) \\
									& 0.13727  	& 0.199(34) & 0.049(15) & 0.300(54) & 0.064(21) & 1.833(144) & -0.098(13) \\
			$\Omega_{c}^0$			& 0.13754  	& 0.320(28) & 0.076(13) & 0.405(44) & 0.096(30) & 1.785(144) & -0.088(10)\\				
			 						& 0.13770  	& 0.313(36) & 0.061(10) & 0.405(38) & 0.053(18) & 1.838(183) & -0.099(18)\\								
			\hline
									& Lin. Fit 	& 0.330(32) & 0.064(10) & 0.398(44) & 0.056(19) & 1.710(150) & -0.099(14)\\
									& Quad. Fit & 0.398(52) & 0.069(22) & 0.484(70) & 0.054(38) & 1.915(279) & -0.083(28)\\														
			\hline \hline
			  						& 0.13700  	& 0.249(29) & 0.071(7) & 0.405(49) & 0.077(11) & -0.402(32) & 0.412(17)\\
									& 0.13727  	& 0.198(18) & 0.051(6) & 0.253(25) & 0.073(9) & -0.356(19) & 0.411(11) \\
			$\Omega_{cc}^+$			& 0.13754  	& 0.276(22) & 0.082(6) & 0.367(40) & 0.096(9) & -0.370(27) & 0.432(9) \\				
			 						& 0.13770  	& 0.316(48) & 0.074(9) & 0.385(47) & 0.088(10) & -0.393(33) & 0.436(15)\\				
			\hline
									& Lin. Fit 	& 0.287(31) & 0.078(7) & 0.350(44) & 0.095(9) & -0.370(26) & 0.441(12)\\
									& Quad. Fit & 0.422(51) & 0.104(13) & 0.534(72) & 0.101(16) & -0.428(58) & 0.453(22)\\														
	\hline\hline
\end{tabular*}
	\label{res_table_quark}
	}
\end{center}
\end{table*}


\begin{table*}[ht]
	\caption{ Goodness of fit analysis of our chiral extrapolations for different fit forms that we use. }
\begin{center}
\begin{tabular*}{1.0\textwidth}{@{\extracolsep{\fill}}cccccccccccccc}
		\hline\hline 
				 & Fit Form & $\langle r_{E,\Sigma^{++}_{c}}^2 \rangle$   & $\langle r_{E,\Omega^{+}_{cc}}^2 \rangle$ & $\langle r_{M,\Sigma^{0}_{c}}^2 \rangle$ & $\langle r_{M,\Sigma^{++}_{c}}^2 \rangle$ & $\langle r_{M,\Omega^{0}_{c}}^2 \rangle$ & $\langle r_{M,\Omega^{+}_{cc}}^2 \rangle$ & $\mu_{\Sigma^{0}_{c}}$ & $\mu_{\Sigma^{++}_{c}}$ & $\mu_{\Omega^{0}_{c}}$ & $\mu_{\Omega^{+}_{cc}}$\\
				\hline \hline
				$\chi^2/$d.o.f. & lin. & 0.916 & 3.701 & 2.265 & 1.823 & 1.560 & 1.162 & 0.440 & 0.993 & 0.129 & 0.149 &\\
								& quad. & 0.222 & 5.988 & 0.003 & 0.065 & 1.624 & 1.804 & 0.079 & 0.005 & 0.098 & 0.139 &\\
				\hline				
				p-Val 			& lin. & 0.40 & 0.025 & 0.10 & 0.27 & 0.16 & 0.31 & 0.40 & 0.64 & 0.37 & 0.86 &\\
								& quad. & 0.64 & 0.014 & 0.96 & 0.69 & 0.80 & 0.18 & 0.70 & 0.78 & 0.95 & 0.71 &\\
		\hline\hline
	\end{tabular*}
		\label{gof}
	\end{center}
	\end{table*}
	

\bibliography{./sbaryons}

\end{document}